# Improving Capstone Research Projects: Using Computational Thinking to Provide Choice and Structured Active Learning

Graham Wild[1]

[1] School of Engineering and Information Technology, UNSW Canberra, ADFA, ACT, Australia

**ABSTRACT:** This work presents a structured systematic process for undergraduate capstone research projects embodying computational thinking (CT) practices. Students learn to conduct research with a decision support system utilizing CT. The system is demonstrated through a case study of a capstone research project course. The course is a 3$^{rd}$ year single semester capstone in an aviation program. CT was integrated over a decade, through 21 semesters of coordinating and delivering the course. The CT practices evolved and were utilized for more aspects over time. The CT system facilitated a significant reduction in staff workload by eliminating the need for direct one-on-one supervision and enabling the streamlining of marking. This resulted in fairer marking by eliminating supervisor bias. Student feedback shows a high degree of satisfaction, with comments highlighting choice and learning.

## 1. Introduction

Architecturally, the capstone is the masonry above an arch, supported by the structural elements beneath it, from the cornerstone to the keystone. The capstone project in higher education fulfils the same role, it is not a new part of the supporting structure, it is above it. In a capstone course, students utilize their knowledge and skills constructed throughout their program in a "substantial real world project" (Herbert, 2018). Capstone projects can help to prepare students for their career by involving real experiences relevant to their profession (Neyem et al., 2014). In the context of undergraduate US institutions, estimates suggest capstones are utilized in 40 to 98 percent of programs, depending on the institute and discipline (Hauhart & Grahe, 2014), which is a rapid increase from three percent in the 1970s. Clearly, the capstone has become a significant aspect of modern higher education.

Many problems associated with capstone courses have been identified in the literature. A potential issue is the significant staff workload required, which is across multiple roles, advisors, instructors, and markers (Novitzki, 2001). Two other issues include generating the required number of topics and student performance (Novitzki, 2001). Farrell et al. (2012) suggest that there are specific challenges when it comes to assessing capstone projects. There is also the issue of fairness when different supervisors and markers are involved. Lawson et al. (2015) states that the course coordinator is the key to achieving fairness, by ensuring the course and the associated assessments are designed to support supervisors, markers, and any necessary moderation. The key issue identified by Lawson et al. (2015) was supervisor bias, which typically requires additional workload for mark adjustment and moderation. Alternatively, supervisors can be removed from the assessment process; although, they have knowledge of the student's work that is not reflected in certain assessed items. While the thesis is a key element to reporting the findings of science, students who are more practically or even mathematically minded may not be fairly assessed with a written thesis or oral presentation.

While groupwork and teamwork activities have benefits, there are also issues. Hassanien (2006) found that "poor communication" and "poor attendance at group meetings" were the primary issues identified by students. Lee et al. (2016) showed high response rates for issues such as personality clashes, which resulted in difficulty compromising. Similarly, different personalities resulted in different levels of engagement and motivation. Another commonly discussed issue is uneven distribution of workloads (Wilson et al., 2018), which can be the result of initial task delegation or devolving situations over the project duration. Finally, LaBeouf et al. (2016) found that the primary issue was around grade assignment and fairness. Again, there are significant benefits that come with groupwork, especially the transference of teamwork skills to the workplace, where working in groups is almost unavoidable. However, another interesting feature to note is that higher-degrees-by-research is exclusively the domain of individual work.

Another complexity of capstone projects is the question of what constitutes a suitable or applicable project, which can vary by disciplines and traditions. As self-directed learning courses (Blanford et al., 2020), two broad categories can be used to describe the types of work undertaken; these are akin to Problem-Based Learning (PBL) and Inquiry-Based Learning (IBL) (Oğuz-Ünver & Arabacioğlu, 2011). That is, students can set out to solve a problem (PBL) or answer a question (IBL); more technical disciplines favor the problem-solving approach while pure sciences favor inquiry. However, variations can come about due to differences in preferences by supervisors or those proposing/providing topics, including the students. This is further compounded considering what is and is not research. Leedy and Ormrod (2013) note that a literature review is not in itself research; however, these are readily proposed topics.

Wing (2006) defined computational thinking (CT) as solving discipline independent problems utilizing the fundamental concepts in computing. Wing (2008) stated that computing in this context referred to "computer science, computer engineering, communications, information science and information technology." These same algorithmic traditions span most of STEM, in particular physics, mathematics, and engineering, where computing solutions is a regular occurrence. This should not detract from the refinements that "pure computer science" has made, but reflects the broader sense that computing is wide reaching. This can be seen in the modern k-12 educational initiatives (Wing & Stanzione, 2016), where "coding" can be applied to storytelling. Lee and Kang (2019) state that the effectiveness of CT applied outside engineering and more broadly STEM specific courses is limited. Their work investigated the application of CT to capstone projects for non-engineering students, which highlighted positive outcomes.

The intent of this Section has been to highlight the capstone project, the importance of such an endeavor, and the associated issues that can arise with them. The aim of this work is to show how CT applied to the design and delivery of a capstone project course can overcome associated issues. To illustrate this, a case study spanning a decade with a capstone project taught over 20 semesters is presented. The structured systematic capstone research project approach was initially envisioned as a "choose your own adventure" capstone project (Mundy & Consoli, 2013), balancing choice, learning, and effort. The evolution of the delivery, dealing with issues, and the improvements in both student satisfaction and outcomes are presented. The discussion highlights how course delivery problems were solved utilizing the practices of CT (Lyon & Magana, 2021).

## 2. The Aviation Capstone Project Case

### 2.1. Higher Education Program Features

Aviation programs can incorporate many aspects associated with business programs, including accounting, finance, management, and marketing. These are specifically contextualized to the aviation industry for airlines, airports, government agencies or bodies, and consulting businesses. Traditional aeronautics can also be included, specifically aimed at students wishing for a degree relevant to pilot training. Further technical aspects of aeronautics can be included for aircraft maintenance and other facets of the industry. Finally, non-technical skills are included covering essential aspects of the industry such as human factors and safety management. Clearly, aviation in higher education has a diverse knowledge base with many potential roles, all within a single narrow industry. The result is that a capstone project in aviation needs to cover a wide variety of potential topics.

Reporting culture is a central aspect of the aviation industry, and is key to achieving high safety standards in airlines (Ulfvengren, 2007). The business aspects of aviation also focus on reporting (García-Sánchez et al., 2013). If an aviation program focuses on the flying, safety, and management aspects, reporting will be a common theme, and hence report writing is a critical skill for students to master. It is not uncommon for a university aviation course to include an individual report assignment and a larger group report, typically with an oral presentation. This addresses practical work-ready skills developed by students which are essential for industry success (Chia & Round, 2015). Nathan (2013) outlines the different forms of case reports representing most reporting activities aviation students complete. However, the case report is located at the bottom of the hierarchy of evidence (Brighton et al., 2003); with students mostly focused on reporting throughout their studies, climbing to higher levels of the hierarchy in a capstone project can be difficult.

## 2.2. The Researcher

It is essential to consider my philosophical orientation toward research (Leedy & Ormrod, 2013), and more importantly how this has evolved. The relevant chronology begins in 2012, as a physicist and mathematician teaching aviation systems and technology. I mostly looked through a positivistic lens and was completely objectivistic, while noting life was never truly black or white. For physics or engineering research, I was applying objectivism and positivism and as an educator I was moderating this with subjectivism and post-positivism. In both scenarios, I believed in realism. I completed a 3$^{rd}$ year capstone project, an honors degree, and a PhD in physics, each with an associated thesis; this gave me a black and white view of research, with research only using filtered information, and unfiltered information used in reporting (analogous to journalism). Now, my research has expanded beyond objective laboratory experiments; I code qualitative data from content analyses and case studies. This required professional development in qualitative methodologies, and I joined the University's Qualitative Interest Group. While typically one should first consider their philosophical orientation to inform their methodology, I believe the reverse is also possible. That is, you can select a methodology and ensure that you orient yourself philosophically to undertake that research, being a methodological chameleon (Martin, 1990).

## 2.3. Case History

In 2012 I saw confused students expected to advance from their legacy reporting activities to research requiring "innovation" with "novel" aspects, utilizing "the scientific method". Their confusion was understandable; they only had a two-hour introductory lecture in week one and were then expected to go forth and just do research, something they had never really done before. Also, different supervisors had different ideas about what were suitable projects:
- some posed literature reviews, below case reports on the hierarchy of evidence (Glover et al., 2006),
- some just proposed case reports/studies,
- some aimed for higher levels of inquiry-based research, and
- some proposed problem-based topics.

The theses from that semester were mostly "super case reports", understandable given the average student had never done a university experiment, nor engineered solutions to problems, and most did not understand research past the tools used for their case report assignments. I instinctively wanted to emulate the activities of my previous research groups for my "team" of students. I scheduled a weekly session where we could all sit and talk about each other's research. My team of about three grew quickly to a collection of about seven students, where classmates who were likeminded also sat in, seeking guidance. Common issues were highlighted; I spoke to them about literature and what was scholarly and not scholarly, about structure and writing, and even formatting and using software (Word, Excel, and PowerPoint).

I inherited the course in semester one of 2013 and aimed to increase the scientific standard. My team from 2012 showed me that aiming too high would be an issue. The goal was above a case study, or at least a substantial collective case study. Colleagues who had PhDs in business and psychology exposed me to the concept that not everything was either literature reviews or laboratory experiments; obvious in retrospect given my previous physics education research including supervising an honors student. In 2013, the plan was to schedule a weekly workshop, emulating the process from 2012. This worked well, given the first semester cohort was half the size of the second semester cohort. However, the result was more student questions, some I could not answer given I did not know enough about the research traditions of business or much outside STEM. For the second semester I organized a guest presenter to come in each week. This covered literature (presented by me), doing surveys (presented by a colleague), how to do interviews (presented by a former news journalist), the methodology of case studies (presented by an education professor), and other topics. In two years, the course had gone from a single two-hour class in week one, to a six-week lecture schedule covering the basics of research. As the course became standardized and systematized, common technical issues persisted; to address this a one-hour lab per week was added in 2015. This used a different piece of software each week (Google Scholar to find literature, Endnote to manage literature, Word to format and edit a journal article, PowerPoint to make research posters, and two labs on Excel for research data).

In 2017 the course started offshore online, going from two semesters a year, to four! Workloads increased, with day one offshore as soon after January 1$^{st}$ as possible, meaning most staff had just started four weeks of annual leave and were unavailable to supervise. A new issue became cultural differences, and especially the asking of questions in class (Chu & Walters, 2013). The lack of topics from other staff required an examination of past topics to provide more. There were recognizable patterns that could be generalized. When examined objectively side-by-side new topics were being iteratively generated. Heuristically, through trial and error, popular topics became seeds for the next generation of the genetic algorithm. The result was a general potential topic space. This process was described as a "choose your own adventure" approach to topic selection (Mundy & Consoli, 2013). The methodologies that aligned with these topics were apparent. The complex process of identifying a research project and selecting a research methodology had been dimensionally reduced. In each of these paired contexts and methodologies, there was an associated accumulation of knowledge about reliable data types and sources. What was previously a long complex sequence of decision making, had been reduced via abstraction to a single choice; each option had associated essential characteristics.

Growth and diversity of academic staff and a growing ability of students to handle more complicated topics meant some staff were aiming higher. This was driven by a doubling of student numbers through an exchange agreement involving international engineering students completing a double degree with aviation. The first capstone course for these students was in semester one 2017 onshore. The small conclave of correlation grew into the entire domain of predictive methodologies. While ambition should be commended, some students with better fundamental skills were being tasked with preliminary exploratory research to assess the potential for future research. If unfruitful, students were left with little to show; this required a pivot, maintaining the context, and simultaneously reducing topic complexity. These pivots were not with the support of supervisors given the distress expressed by students. If possible, these students were encouraged to embrace their negative result, and not focus on the irrelevant goal of positive outcomes. This involved changing research questions to ask if the tools and techniques were applicable or relevant in their aviation context. A more significant issue was if the tools and techniques were above the students' abilities.

To solve the problem of supervisor bias, marking was handled by an objective 3$^{rd}$ party, and this was only facilitated by the increase in personal workload and being provided marking support. Supervisors were invited to mark the weekly reports; an invitation no one accepted. As a result, all but the presentations were marked by an independent marker. The 5-minute proposals (enabling 15 students to present an hour, which was just enough time for 45 students in 3 hours) and the final poster conference presentation were marked collectively usually by 4 to 6 academic staff members and PhD students, with all supervisors invited.

At different times the problem of "what is research" would reappear. Leedy and Ormrod (2013) had been the class text since 2014, selected because the School purchased every staff member a copy. Three key issues were identified. The first was topics that were just literature reviews previously identified as not research. Second was not applying methods appropriately like using qualitative analysis for quantitative research. Finally, there were supervisors who just wanted another case report, which no longer aligned with the course learning outcomes. To try and resolve these issues the offshore online topic structure was adopted in 2018. Students were presented with the option to use the "choose your own adventure" or engage with an academic staff member to work on a specific topic. Although, growing postgraduate numbers reduced the number of unique topics from other academic staff. Students were cautioned, supervisors were not responsible for marking, they did not set the course learning outcomes, and while there may be different opinions on what was research, there were very clear minimum standards thoroughly documented for the course. Each issue was discussed with students in week one when talking about topic selection, and how to manage them. If a supervisor wanted a literature review, the student used content analysis with the literature as data following a well-defined methodology (Snyder, 2019). Most students adopted the "choose your own adventure" style, and the main reason not to was because they had a specific desire to work with a particular member of staff, enabling them to engage in the student-staff partnership of their choice, providing value (Martens et al., 2019).

## 3. Computational Thinking for the Capstone Research Process

Figure 1 shows the original process from 2012. The process started with the "context" (in 2012 this term was topic). The context is the knowledge domain or aspect of aviation the student will focus on. This could be airline operation or aircraft maintenance, etc. This choice is the first reduction in potential topics. The scoping process was next, and a preliminary version of this came from supervisors with the provided topics. Students would then decide on their specific objectives (research questions), and select their methodology, with reference to the literature and with guidance from a supervisor. These were significant steps that were completed with a lack of knowledge in terms of what the options were.

*Figure 1.* Original project "flow" chart presented to students in 2012

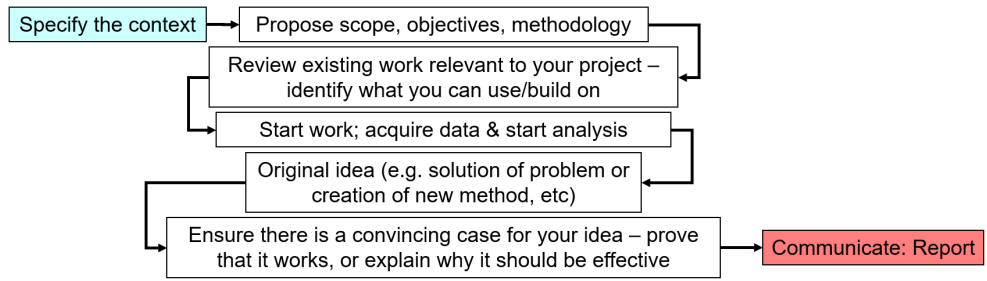

To better define the process of an undergraduate capstone research project, a structured problem decomposition approach was taken to divide the process into algorithmic procedures that the student steps through, inputting some choice or following the automation. That is, the previous project flow evolved over time to a systematic CT inspired process.

**3.1. Research Preliminaries**

*3.1.1. Research Context*

In the current CT inspired approach, students are asked to consider all courses studied prior and score them based on performance and/or interest. Prior performance is important as it is a good indicator of future performance and interest will help provide intrinsic motivation. Table 1 shows the simple nature of this activity, and many students can readily identify their preference (and do not require that activity).

*Table 1.* Program course ranking activity example to determine research context

| Course | Rank |
|---|---|
| Airlines | |
| Airports | |
| ⋮ | |
| Meteorology | |
| Maintenance | |
| ⋮ | |

*3.1.2. Research Philosophy*

A researcher's philosophical orientation towards research is much more important for higher-degrees-by-research. These concepts are complex and definitions are not agreed upon (Bates & Jenkins, 2007). However, three simplified questions can provide conditional logic to determine if the student should undertake qualitative or quantitative research. The three questions are:
1. Ontology (the nature of being): "what can we know, are things real, or are they relative?"
2. Epistemology (the theory of knowledge): "is knowledge a universal truth (objective), or does it depend (subjective)?"
3. Approach (theoretical perspective): "can everything be directly measured (positivistic) or not (post-positivistic)?"

Figure 2 shows the logic circuit representing the decision-making algorithm. If all responses are 1, this suggests quantitative research, with all responses 0 suggesting qualitative. Any nonexclusive configuration could be a mixed method, with a single 0 suggesting QUANT-qual, while any single 1 suggests QUAL-quant.

*Figure 2.* Logic gate circuit to give methodology from research philosophy dimensions

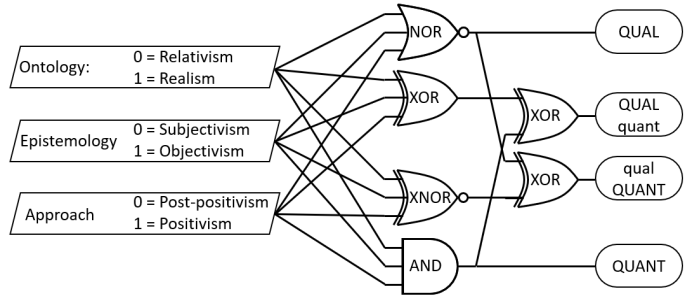

### 3.1.3. Research Subject

The specific subject of research can be thought of in terms of people, records, thoughts, processes, and things. While studying people may involve thoughts and processes, there are some specific guidelines provided:
1. People as a subject implies uncontrolled but active data collection.
2. Thoughts as a subject implies uncontrolled but passive data collection.
3. Processes involving people will be controlled data collection.

The active-passive dimension refers to whether the research or subject is responsible for the data. The controlled-uncontrolled dimension is about stimuli, and if it is random or manipulated by the researcher. Processes can be substantial or insubstantial and include much of the natural sciences (energy and kinetics etc.). Records refers to documents and reports, very common in aviation. Things covers almost anything else, including any secondary data about people or organizations. By selecting one of these five and knowing the output of Figure 2, the student is presented with a suggested research methodology, given in Table 2.

*Table 2.* Research methodology identification based on subject selection

|  | Quantitative | Qualitative |
|---|---|---|
| People | Observational Study | Ethnography |
| Records | Ex-Post-Facto Study | Content Analysis |
| Thoughts | Survey | Phenomenology |
| Process | Experiment | Grounded Theory |
| Things | Developmental Design | Case Study |

### 3.1.4. Research Topic

The ten options from Table 2 prescribe ten "fill in the blank" topic statements (Table 3). The blanks are the subject matter and the context, both of which need to be refined from the general to the specific. For example, people could be passengers, pilots, controllers, cabin crew, ground handlers, engineers, operators, administrators, supervisors, managers, executives etc.

*Table 3.* Ten generic research topics needing subject and context input

| |
|---|
| An observational study of (insert group of people, limited by the context). |
| An ethnographic study of (insert group of people, limited by the context). |
| An ex-post-facto study of (insert type and/or source of records being studied, limited by the context). |
| A content analysis of (insert type and/or source of records being studied, limited by the context). |
| A survey of (insert group of people)'s perceptions of (limited by the context). |
| A phenomenological study of (insert group of people) in (limited by the context). |
| An experimental study of (insert process and relevant details) in the (limited by the context). |
| A grounded theory of (insert process and relevant details) in the (limited by the context) |
| A (longitudinal or cross sectional, select one) study of (describe thing to be studied) in (limited by the context) |
| A case study of (describe thing to be studied) in (limited by the context) |

*3.1.5. Research Questions*

For the primary research question (PRQ) the complexity of the topic needs to be considered; potentially there will be people (who), doing things (what), using a specific method or quantity (how), in a location (where), at a time (when), and for a reason (why). For each research question, the subject of the investigation suggests the correct interrogative word. "Which" will tend to result in a one-word answer and should be avoided. "Why" questions should be avoided given the one semester constraint. Typically, "what" and "how" are used at this level, with "who", "when", and "where" more likely for secondary questions.

Problem decomposition is essential at this point. The resultant PRQ might be overwhelming at the start of the research process assuming it has been well defined. By decomposing the PRQ down, students can avoid feeling overwhelmed. For inquiry-based research, the decomposition is generating secondary research questions (SRQs). There are two research question models, "1<3" where one PRQ is answered utilizing three SRQs, and "1+2" where there is one direct PRQ and two related SRQs. If a student needs to answer more fundamental questions before they can answer their PRQ, these will be their SRQs; answering them will help to (or directly) answer the PRQ. Alternatively, if there is a direct procedure between the student and the answer to their PRQ (they are not overwhelmed), then they pose some less important parallel questions that can be answered with little extra effort. These will be their SRQs.

Pseudo questions also need to be discussed, as these are common for first time research students. They may not know how to measure their subject, so they pose this as an SRQ. However, "how to measure" is just a procedural step, requiring new knowledge, not research. These aspects are included in the section on research methods. Another common pseudo question posed by students is around economic impact, that is "how much does the phenomena cost the industry every year?" The only way to prevent this was to require all students to include this as part of their research significance and impact in the introduction.

Finally, students need to pose hypotheses for their research questions. As a capstone, students have their prior knowledge from their program/degree. Therefore, they should be able to conjecture potential answers for their research questions. The hypothesis is another hurdle students can have difficulty getting over. They are usually worried about their hypothesis being wrong. Hence, students are told that their hypothesis being wrong is beneficial because it generates more discussion at the end of the research, enriching their discussion section. This is simply meant to help students not fixate, and they are told a concerted effort should be used to consider their hypotheses fully.

*3.1.5. The Overall Preliminary Process*

The overall process for the research preliminaries, as described above, is illustrated in Figure 3. The overwhelming unfamiliar prospect of research is now the structured systematic processing of information, taking discipline context in terms of courses previously studied, and generating research questions for a capstone research project.

**3.2. Literature**

The CT inspired literature review process has evolved to control the sources utilized. Ideally only narrow databases should be used (Scopus etc.), noting there may be resource limitations. The quality of the source of literature being used is something the students are taught, prioritizing the academic literature (journals, conferences, edited book chapters, and theses), supported by general and professional sources (books, trade publications, and reports), while avoiding unverified web sources (blogs, forums, etc.). Students are then presented Table 4 indicating where to utilize the different sources. The benefit to using databases like Scopus is that all results will be academic. In Google Scholar, books and patents are included, so too are reports. The point is not to question if an authored book belongs in a literature review, it is to provide simple conditional logic that should prevent confusion and errors. Based on experience, the actual debatable item is the report. While reports from research institutes (NASA, etc.) would be expected in a relevant PhD, in aviation there is an abundance of corporate reports which if used as literature can limit objectivity. As such, this hard line is just to eliminate "type 2 errors" while allowing "type 1 errors". That is, some acceptable sources will be forbidden to ensure no unacceptable sources are allowed.

*Figure 3.* Preliminary research process flow chart, from discipline content to research questions

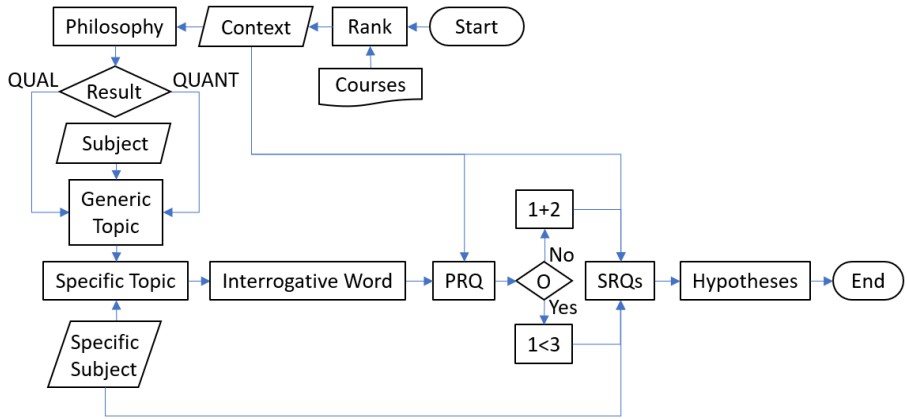

*Table 4.* Usage of different types of information sources in different sections of a thesis/paper

|  | Academic | Supporting | Unverified |
|---|---|---|---|
| Background | Y | Y | N |
| Literature | Y | N | N |
| Methodology | Y | Y | N |
| Discussion | Y | N | N |

Literature is collected at a pace of four items per week, specifically from academic sources. These are presented in one week, then they are summarized into a sentence or two (or a paragraph) in the next week, along with the next four references being identified. This systematic processing of information makes use of iterative, recursive, and parallel CT practices. It decomposes a large literature review (collecting and reviewing approximately 40 academic sources) into SMART weekly goals (Rubin, 2002). The literature review procedure is illustrated in Figure 4, where raw references are "stored" and reported numerically in the order found. When they are reviewed, they are "stored" and reported alphabetically (based on the referencing style utilized). The double handling is a marking feature, where weekly reports require a complete list of literature in the order found; this starts at four in week two becoming 40 in week 11. This was needed to prevent the practice of recycling references previously reported. The predefined processes are covered in laboratory activities (Lab 1 – Endnote, Lab 2 – Literature), and in the lecture program (Week 2, Chapter 3 Review of the Related Literature).

*Figure 4.* The procedural literature review process

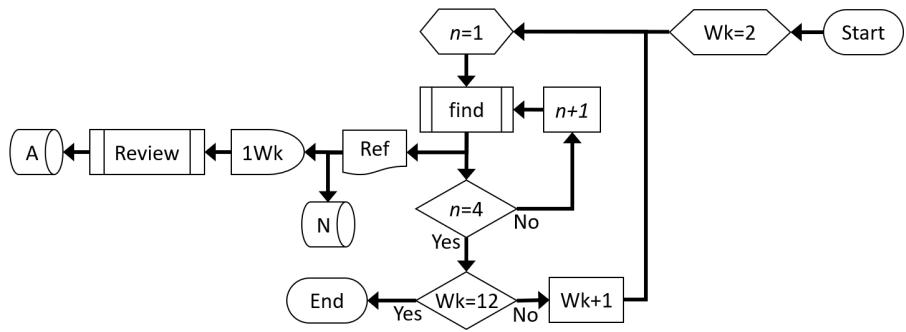

### 3.3. The Introduction

In addition to weekly literature goals, the introduction is also divided into weekly elements, made up of the aim, background, and significance, along with research questions and hypotheses. In week 3 the students provide a draft aim, then in week 4 the background and the significance. During these weeks they are also collecting a glossary of terms, abbreviations, and initialisms. Weekly submission templates are provided identifying the required items, as summarized in Table 5.

Table 5. Weekly breakdown for introduction, literature, and methodology tasks from week 1 to 6

|  |  | Week 1 | Week 2 | Week 3 | Week 4 | Week 5 | Week 6 |
|---|---|---|---|---|---|---|---|
| Introduction | | Context Specific Topic | PRQ SRQs Glossary | Aim Hypotheses Glossary | Background Significance Glossary | - - Glossary | - - Glossary |
| Lit | Find | - | # 1 to 4 | # 5 to 8 | # 9 to 12 | # 13 to 16 | # 17 to 20 |
| | Review | - | - | # 1 to 4 | # 5 to 8 | # 9 to 12 | # 13 to 16 |
| | Other | - | - | - | - | Survey | Themes |
| Methodology | | Quant/Qual | Subject Gantt Chart | - | - | Data Source Procedure | Design Analysis |

### 3.4. The Methodology

During the first six weeks, the students cover 2 chapters a week from Leedy and Ormrod (2013). Week 3 covers the two writing chapters (5 and 13). Week 4 to 6 then cover the chapters on quantitative and qualitative methodologies, as well as mixed methods and analyzing data. The week 5 and 6 submissions require the students to build elements of their methodology (indicated in Table 5), and they are provided with guidance and example papers with an aviation context. A great deal of effort was exerted finding aviation examples on almost every methodology and variation presented in Leedy and Ormrod (2013), as well as every data analysis tool commonly utilized. The student should be able to identify the entire research design applicable to them by week 6, stating if the research will be monomethod, multimethod, or mixed method, and which methodologies and how they are combined. The exact tools to be utilized for data analysis also needs to be identified. This could be stating the required statistical tests to be utilized or providing a start list for qualitative data coding.

### 3.4. Remaining Sections

As with the front "half" of the thesis, the aspects for the back "half" follows a similar decomposition, as summarized in Table 6. This covers data collection, analysis, discussion, and a conclusion. It should be noted that currently, the students have a midsemester break between week 6 and 7, and some data collection is expected and encouraged during this period.

Table 6. Weekly breakdown for literature, data, and discussion tasks, from week 8 to 12

|  |  | Week 7 | Week 8 | Week 9 | Week 10 | Week 11 | Week 12 |
|---|---|---|---|---|---|---|---|
| Lit | Find | # 21 to 24 | # 25 to 28 | # 29 to 32 | # 33 to 36 | # 37 to 40 | - |
| | Review | # 17 to 20 | # 21 to 24 | # 25 to 28 | # 29 to 32 | # 33 to 36 | # 37 to 40 |
| | Data | Collect | Reduce | Analyze | Present | - | |
| Discussion | | - | - | - | Finding Recommendations | Assumption Limitation | Future Work Conclusion |

## 4. Student Feedback

In 2020 the CT based capstone research course was implemented at a new university. This was necessary given staff members had just retired or moved to industry, and of the three new staff, one was starting in 2021, and the other in 2022. As a result, there was only one staff member to supervise students in 2020, and given COVID-19 restrictions, the previous online offshore version of the course was utilized. This presented an ideal opportunity to test the CT process by requiring students generate a project using the structured systematic capstone research project procedure. While the quantitative scores for student feedback are impressive (100% agree overall for course and lecturer for both years) the small number of students limits analysis to qualitative data.

The key themes identified in the responses addressing the courses "best elements" were:
1. Choice
2. Learning (prepared, trained),
3. Structured (weekly), and
4. Examples

The quote "[t]he broad scope of what our topics could be" as the "best element" highlights that the goal of choice has been achieved. Looking at this from the staff perspective, the students are "limited" to do one of 10 generic topics, which only requires a finite knowledge base for the coordinator, lecturer, and/or marker; however, the students perceive almost limitless possibilities. This is critical to the success of the CT approach when it is combined with item 2, learning. That is, every student is taught the basic elements of all 10 generic topics, including all the different data analysis tools utilized. The significant quote for learning is:

> [The best element was] [a]ll the structure and resources. While the onus was on us to do our own research and take the project where we wanted to go, we actually learned the research theory first/concurrently, so we had something to go on! It would be impossible to expect us to just know how to research and write reports and lit reviews properly without some sort of education (which to be fair is what Advanced topics in [removed] and [removed] is like).

For context, several students were simultaneously doing their capstone in aviation and for their other science major. The point emphasized by the student is that they felt empowered by the learning and noted the familiar legacy approach in other capstone courses. Clearly this student is not doing a triple major, so for them to be able to note two other disciplines suggests that the students discussed this specific point.

CT practices have also been implemented in other courses in terms of marking and feedback (Author, 2018). This was developed extensively in a first-year course taken by 150 science and engineering students in aviation and aerospace. For the engineering student the course helped achieve the Engineers Australia graduate attribute concerning "effective (…) written communication in professional and lay domains" (Kootsookos et al., 2017). Ensuring effective written communication could be assessed (and was substantial) with 150 students necessitated an algorithmic approach. These principles were carried over into the marking and feedback for the capstone project course, where there were up to 55 aviation students in a single semester class. The process of marking utilized Google (or Microsoft) Forms that captures the rubric elements, scoring points, penalties (regarding weekly literature item quality), and common feedback statements. The goal is to facilitate efficient and effective feedback.

The only interesting feature of the "worst element" was about more feedback. The reason this was flagged (by one student in 2021) is likely that "corrective feedback" is used, and only in a negative sense (Cheung & Molesworth, 2022). That is, the described CT inspired systematic iterative marking procedure is only designed to tell students what they have done wrong. In principle if a student made no mistakes, but was unsure, they would see no substantial comments (other than a programmed "well done"). The proposed solution here is to add a self-reflection section to the weekly reporting, so students can indicate if they felt less confident about a specific section and to ask questions; additional comments could be added in the feedback to address these.

## 5. Discussion

This case study has shown that a capstone course can run effectively as a conventional course, with a single academic staff member (with marking support depending on class size), previously demonstrated with limitations (Esselstein, 2013). The reflexive thinking detailed in Section 2.3 and the qualitative results of the student feedback highlight the need to include teaching, which combined with the structured systematic capstone research project procedure facilitates choice resulting in positive outcomes. The approach is an improved "scalable and portable structure for conducting successful" capstone projects (Keogh et al., 2007), in any academic discipline, and potentially even for full year and/or forth year courses, as well as for coursework master's programs where time is limited (Izu, 2018).

Choice was a key factor guiding change, given choice promotes enjoyment and variety (McBride, 2009). Choice is also directly linked to motivation (Rodriguez et al., 2017), and is important to capstone projects (Kaela et al., 2019). That said, not all students want to "choose their own adventure", some want to do unique things, and still others want things prescribed. That is, some students want to choose not to choose (Sunstein, 2014). The structured systematic capstone research project procedure respects the choice by replacing "unlimited" choice with several objective questions, and a choice between five options, the subject, which can be defaulted to "thing". However, for those who want to choose, the system provides options relevant to their discipline interests, the number of choices is bounded, and the choices are not too complex. Katz and Assor (2007) state choices like these "enhance motivation, learning, and well-being." The provided choice through the applied CT practices may result in students with either an internal or an external locus of control being satisfied (Harrison et al., 1984).

Issues have been identified in engineering capstone courses in Australia (Rasul et al., 2015). These include lack of preparation, serious marking concerns (including supervisor bias), what was being assessed (the journey, the outcome, or the thesis), and limited supervisor training and support. All four of these issues are resolved with the structured systematic capstone research project procedure. Preparation is covered with the teaching of research methods. Marking concerns would still exist, and moderation would be needed for very large cohorts; however, this would be for objective 3$^{rd}$ party markers, and a suitable 10% overlap could address this. The serious concern of supervisor bias has been resolved. Finally, course coordinator training and support is necessary. As discussed in Section 2.2, training and professional development in the array of methodologies is helpful. However, the large-scale lack of training and support for dozens of supervisors is eliminated.

The CT inspired capstone research project decomposition has helped to resolve issues noted with the use of remote teaching and learning (Clear, 2011). That is, the approach can be used for offshore teaching, demonstrated with six successful consecutive semesters offshore, using only one academic staff member. COVID-19 has also impacted traditional university education, and will have an ongoing effect (Webb et al., 2021). As such, the ability to offer effective and efficient capstone research courses will become more important. There is an associated limitation, and that is the generation of experimental data. While there are potential opportunities for some at home experiments, this would be very limited. Remote simulation and modelling topics would be common.

Since the structured systematic capstone research project procedure was intended to address inquiry-based research, there may be a limitation for PBL capstone courses. Looking at the literature on capstone projects, there are many different approaches (Rasul et al., 2015). IT capstone projects can be very focused on the development and delivery of an IT solution (Izu, 2018). Product design and prototyping are also common aspects of engineering capstone projects (Higbee & Miller, 2020). It has been noted that research aspects can be integrated into IT capstone projects (Chard & Lloyd, 2014), and the identified methodologies could be applicable to engineering capstone research projects. A possible solution could be two capstone courses, each of a semester, one for project or product design and development, and the second on research adopting the product/approach designed.

A lack of teams/groups may be considered a limitation. In the context of aviation, the consistent utilization of groupwork in previous courses addresses this. However, teamwork could be utilized by grouping students according to their methodology. While not a group project, students with related specific topics could work together through the research process. However, a comparative study between the individual student's subjects could be a groupwork extension. Mentoring of these teams by additional academic staff members would also be an interesting practice.

## 6. Conclusion

The notion of "choose your own adventure" was the original inspiration for the capstone research procedure. In these novels from the 1980s, there were branching storylines in a single book, and you could choose to turn left or right, turning to the corresponding page to continue the story. This resulted in different outcomes for different readers, with the same page limit. The key point here is that the incorporation of CT practices does not mean the lack of variety and potential, nor does it mean a lack of creativity, imagination, and freedom; rather, the algorithms and automation provide a scaffold which countless capstone projects can be built with, just as the tree search that makes up a "choose your own adventure" can still facilitate artistic creativity (Newman, 1988). Interestingly, there is a body of literature on "choose your own adventure" in education which will also retrospectively embody CT practices.